\documentclass[aps,pra,notitlepage,twocolumn,superscriptaddress]{revtex4-1}
\usepackage{amsmath,amssymb,amsfonts,bm,comment}
\usepackage{graphics,float,tikz}
\usepackage[caption=false]{subfig}
\usepackage[colorlinks=true,
	linkcolor=blue,
	filecolor=magenta,      
	urlcolor=blue,
	citecolor=blue]{hyperref}
\usepackage{physics,xparse,mathtools,isotope}
\usepackage[table-parse-only]{siunitx}
\usepackage{multirow,array,relsize,lipsum,longtable,tabularx}
\newcommand{\im}{\mathrm{i}}

\begin{document}
	\title{Transfer of orbital angular momentum superposition from asymmetric Laguerre-Gaussian beam to Bose-Einstein Condensate}

\author{Subrata Das}
\email{subratappt@iitkgp.ac.in}
\affiliation{Department of Physics, Indian Institute of Technology Kharagpur, India}
\author{Anal Bhowmik}
\affiliation{Department of Mathematics, University of Haifa, Haifa, Israel}
\affiliation{Haifa Research Center for Theoretical Physics and Astrophysics, University of Haifa, Haifa, Israel}
\author{Koushik Mukherjee}
\affiliation{Department of Physics, Indian Institute of Technology Kharagpur, India}
\author{Sonjoy Majumder}
\email{sonjoym@phy.iitkgp.ac.in}
\affiliation{Department of Physics, Indian Institute of Technology Kharagpur, India}
\date{\today}

\begin{abstract}
In this paper, we have formulated a theory for the microscopic interaction of the asymmetric Laguerre-Gaussian (aLG) beam with the atomic Bose-Einstein condensate (BEC) in a harmonic trap. Here the asymmetry is introduced to an LG beam considering a complex-valued shift in the Cartesian plane keeping the axis of the beam and its vortex states co-axial to the trap axis of the BEC. 
Due to the inclusion of the asymmetric nature, multiple quantized circulations are generated in the beam. We show how these quantized circulations are transferred to the BEC resulting in a superposition of matter vortex states.
The calculated Rabi frequencies for the dipole as well as quadrupole transitions during the transfer process show distinct variability with the shift parameters of the beam.  A significant enhancement of the quadrupole Rabi frequency for higher vorticity states is observed compared to symmetric single orbital angular momentum (OAM) mode beam at a particular range of the shift parameters. We also demonstrate the variation of superposition of matter vortex states and observe its distinct feature compared to the superposition of the LG modes for different shift parameters. The first order spatial correlation of the superposed states supports this feature and highlights asymmetry in degree of transverse coherence along orthogonal directions on the surface.

\end{abstract}
\maketitle
\section{Introduction}

Coherent quantum superposition of vortices \cite{bhowmik_2016_interaction, bhowmik_2018_tuning, kanamoto_2011_quantum, thanvanthri_2008_arbitrary, hallwood_2010_robust}   in atomic Bose-Einstein condensate (BEC) using external field has proved immense potentiality in technology \cite{gullo_2010_vortex, kapale_2005_vortex}, especially in the area of interference using vortex-antivortex pair \cite{liu_2006_struc, wen_2008_dynamic, wen_2013_struc_two}. There are other recent frontier engineerings with vortices at BEC such as vortex nucleation \cite{price_2016_vortex},  fractional quantum circulation \cite{kanai_2018_flows}. 
After the pioneering work of Allen \emph{et al.} \cite{allen_1992_lg} on optical vortex carrying orbital angular momentum (OAM), associated with its spatial mode structure, there have been remarkable advancement in creation \cite{agarwal_1997_vortex,arlt_1998_produc,sueda_2004_laguerre}, manipulation \cite{akamatsu_2003_coherent} and detection \cite{molina_2001_manage,bigelow_2004_breakup,leach_2002_measure} of the OAM states of light along with its utilization to generate the vortex states in BEC \cite{andersen_2006_quant}. In this regard,  the utilization of optical vortex carrying OAM is already established as an attractive opportunity in high-density data transmission \cite{gibson_2004_free}, manipulating the motion of microparticles in optical
tweezers \cite{he_1995_direct}, optical trapping of atoms \cite{kuga_1997_novel,otsu_2014_direct,kiselev_2016_optical,kennedy_2014_confine,bhowmik_2018_tunable} and quantum information processing \cite{mair_2001_entang,molina_2001_manage,zou_2005_scheme,giovannini_2011_resilience,garcia_2012_quantum}. However, the interaction of an atom with optical vortex, in the dipole approximation, inevitably transfers OAM to the center-of-mass (c.m.) of the atom, below the recoil limit, and rotates the atom around the axis of the beam \cite{vanEnk_1994_selection,vanEnk_1994_commu,babikar_2002_oam,alexandrescu_2005_elec,jauregui_2004_rota,alexandrescu_2006_mecha,wright_2008_optical}. This transfer  mechanism generates quantized vortices in the atomic BEC either through Raman processes \cite{nandi_2004_vortex,andersen_2006_quant,mondal_2014_oam,bhowmik_2018_density} or slow light technique \cite{dutton_2004_transfer}. Moreover, only one recent theoretical work \cite{mondal_2014_oam} followed by two experimental realizations \cite{schmiegelow_2016_tran,giammanco_2017_influ} demonstrate the detail picture of the transfer procedure of the OAM from the optical vortex to the electronic motion of the atom. Furthermore, such transfer process has been used to realize  vortex-antivortex superpositions, whose applications are well studied in the literature \cite{gullo_2010_vortex,kapale_2005_vortex,groszek_2018_vortex}.

Propagation of off-axis Gaussian and higher order modes of a light beam in lens-like media with spatial gain or loss variation leads to astigmatism and asymmetry in the beam \cite{tovar_1991_off,al_rashed_1995_decen}.
Different asymmetric properties of light beams propagating through the turbulent atmosphere have been investigated in recent years \cite{cang_2013_propa,zhu_2008_cohe}. Vasnetsov et al. \cite{vasnetsov_2005_analysis}  showed that misaligned Laguerre-Gaussian (LG) beam can be represented as superposition of Bessel-Gaussian beams carrying well defined OAM. Phase structure and intensity of the LG beam can be manipulated by coaxial superposition of the beam with the help of off-axis hologram \cite{vaziri_2002_super,ando_2010_struc,parisi_2014_mani}. Kovalev \emph{et~al.} \cite{kovalev_2016_alg} introduced asymmetric Laguerre-Gaussian (aLG) beams generating  OAM  non-linearly depending on the asymmetry parameter and produced the crescent-shaped intensity beam pattern. This modified OAM provides extra degree of freedom for two-photon entanglement if aLG beam is used as pumping laser beam \cite{mair_2001_entang,kovalev_2016_alg}. However, to the best our knowledge, no investigation on the interaction of aLG beam with the ultracold atoms, in particular, atomic BEC has been addressed in the literature so far. \\
Motivated by the limited number of literature on the interaction of such singular light beam \cite{dennis_2009_singular,desyatnikov_2005_optical} with matter-waves 
  compared to the richness of the phenomenon, here, we investigate the microscopic interaction of the aLG beam with the ultra-cold atoms and employ the developed formalism to generate a superposition of vortex states in the latter. Since the aLG beam can be expressed as a summation of co-axial LG beams with multiple quantized OAM (see Eq. \eqref{eq:alg_exp}), the interaction of the aLG  beam with BEC is expected to produce a superposition of multiple vortices in the latter.
One of our recent works \cite{bhowmik_2016_interaction} showed that multiple vortices in BEC can be created even from single  LG beam using two-photon Raman transition if the beam is non-paraxial. In that case, the creation of the superposed state is generated via three different intermediate electronic states due to the spin-orbit coupling of light and also the superposition is very weak. However, for an aLG beam, two-photon Raman transition requires only  single intermediate electronic  state and we observe superposition multiple vorticities with the same sign of orientation.

Long range of spatial ordering of coherence is well known for non-vortex BEC \cite{andrews_1997_observation, bloch_2000_measure}. However, our analysis here  along the two dimensional cross-section of the condensate shows smaller range ordering of the vortex system and which even varies non-uniformly in different directions on the cross-section with the shift parameter of the aLG beam. This feature is consistent with the tomography of density.
Therefore, this is an unique mechanism  in ultra cold atoms similar to condensed matter systems such as superfluids, type-II superconductors, quantum-Hall effect materials, and multicomponent superconductivity \cite{engels_2002_nonequi,paredes_2001_anyons,paredes_2002_fermion,abo-shaeer_2001_observation,ho_2001_bose,fischer_2003_vortex,anders_2005_fractional,milorad_2015_emergent}.

The paper is organized as follows. In Sec. \ref{sec:theory}, we discuss the properties of the aLG beam and formulate the corresponding interaction theory between the aLG beam and cold atom. In this section, we have also proposed a model to create a superposition of vortex states in BEC using single  aLG beam. Section \ref{sec:res_dis} describes the numerical results and discusses the superposition of final states which can lead to a large quantum number entanglement as suggested by Fickler et al. \cite{fickler_2016_quantum}.  In the last section,  we conclude our results and present some theoretical as well as experimental outlooks of the paper.

\section{Theory}\label{sec:theory}
\subsection{Asymmetric Laguerre-Gaussian beam}
\begin{figure}[H]
	\centering
	\subfloat[][$\delta=0,\beta=0$]{\includegraphics[width=0.23\textwidth]{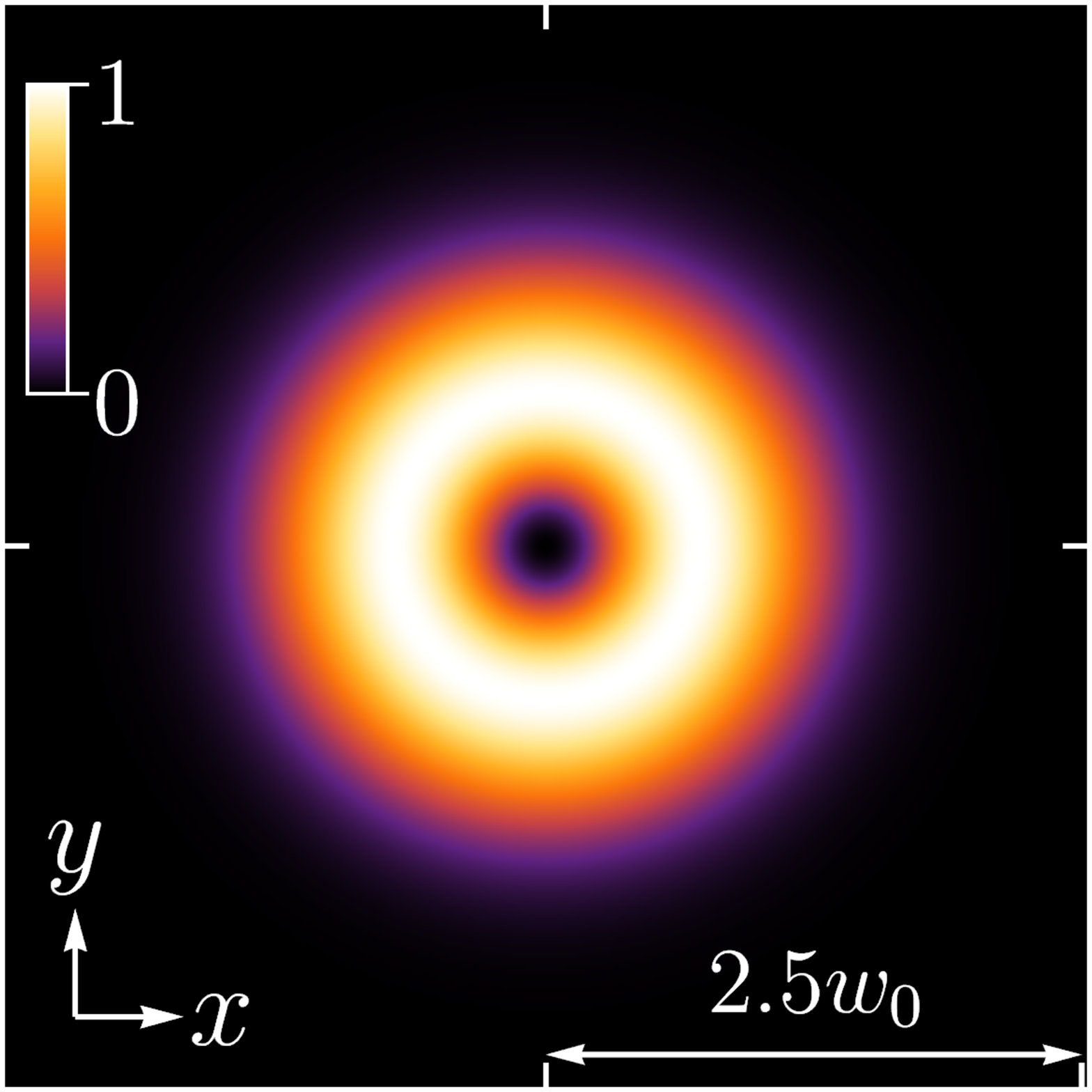}
		\label{fig:alg_z0_0_P0}}
	\subfloat[][$\delta=0.25w_0,\beta=0$] {\includegraphics[width=0.23\textwidth]{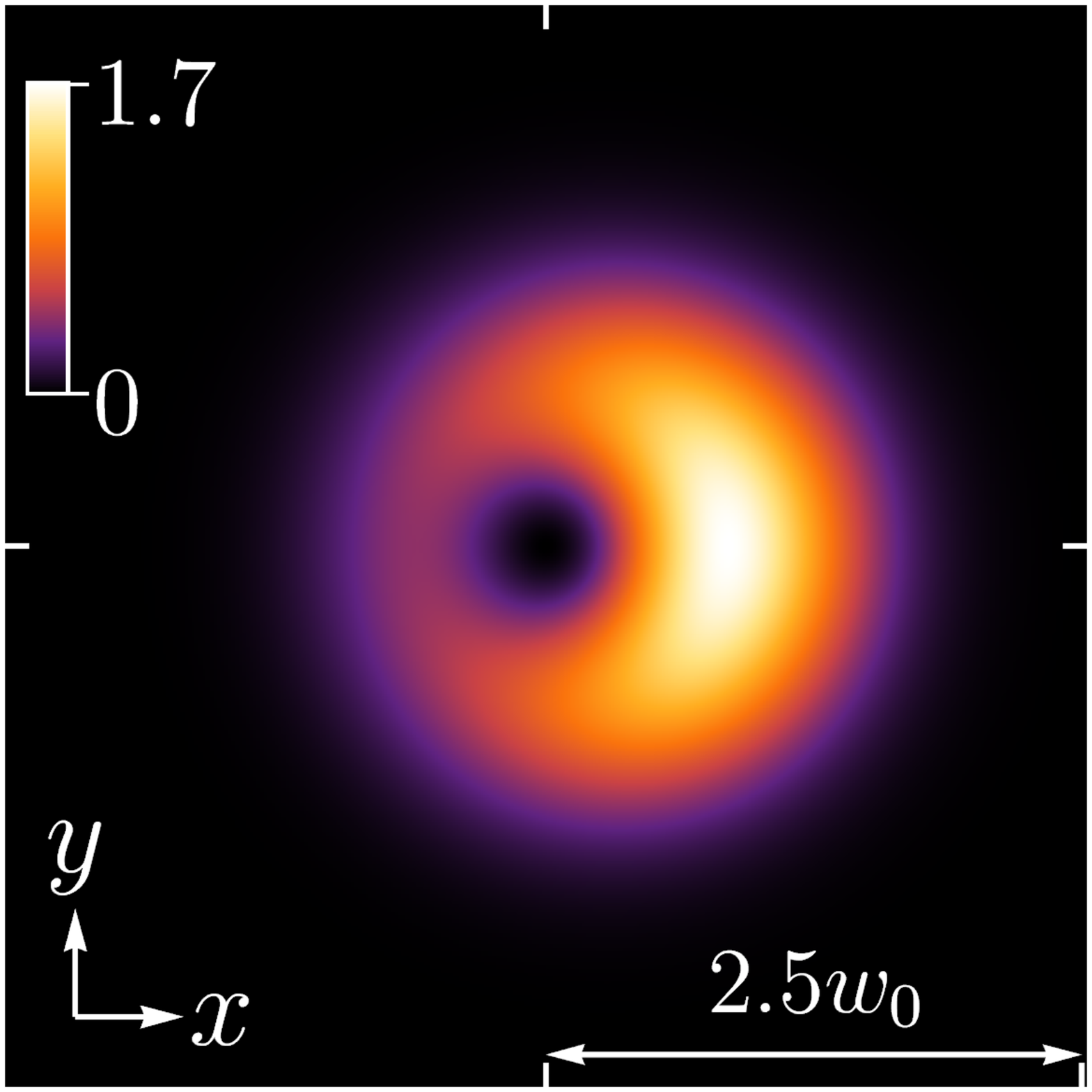}
		\label{fig:alg_z0_25_P0}} \vspace{0cm}
	\subfloat[][$\delta=0.5w_0,\beta=\frac{\pi}{4}$]{\includegraphics[width=0.23\textwidth]{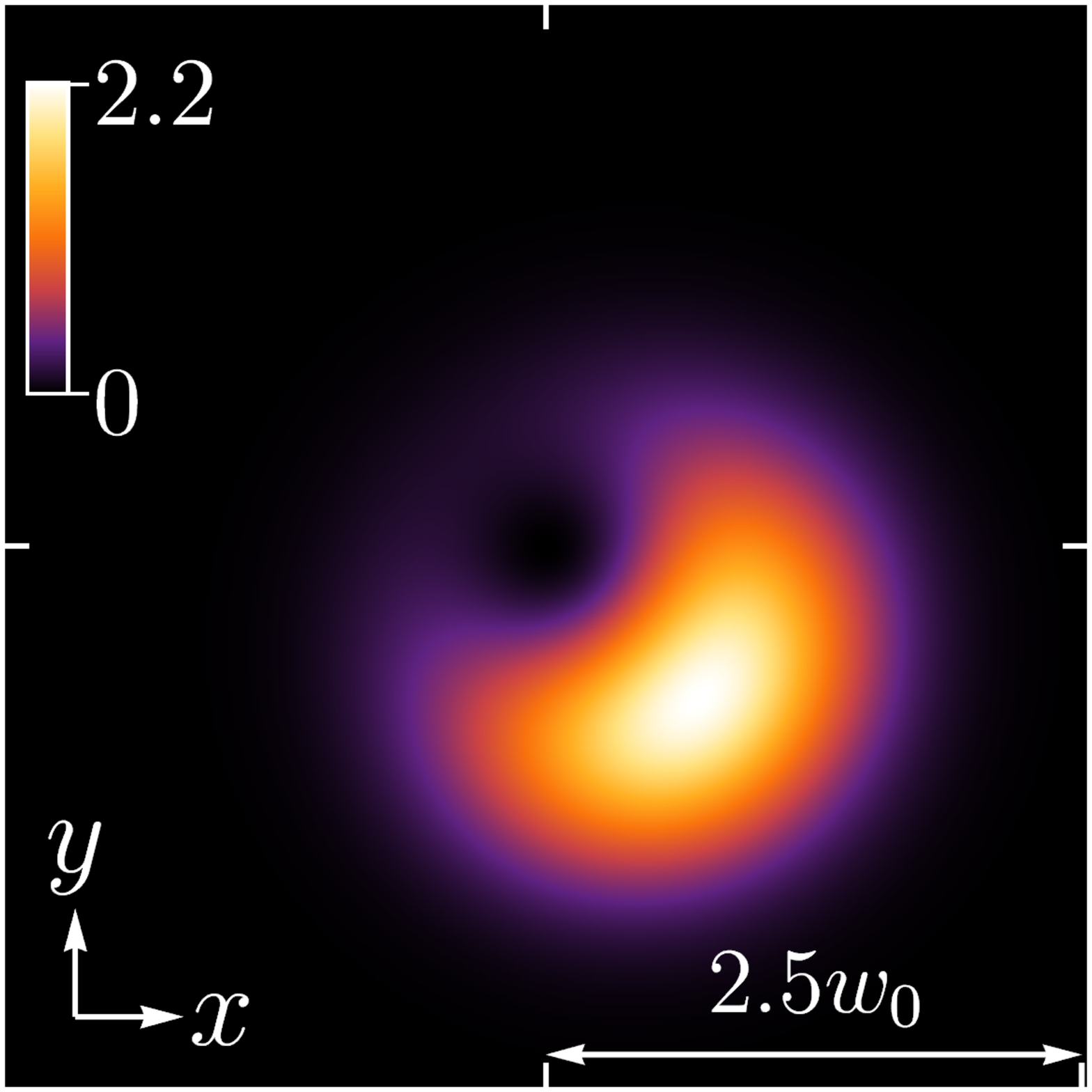}
		\label{fig:alg_z0_50_P45}}
	\subfloat[][$\delta=0.75w_0,\beta=-\frac{\pi}{4}$]{\includegraphics[width=0.23\textwidth]{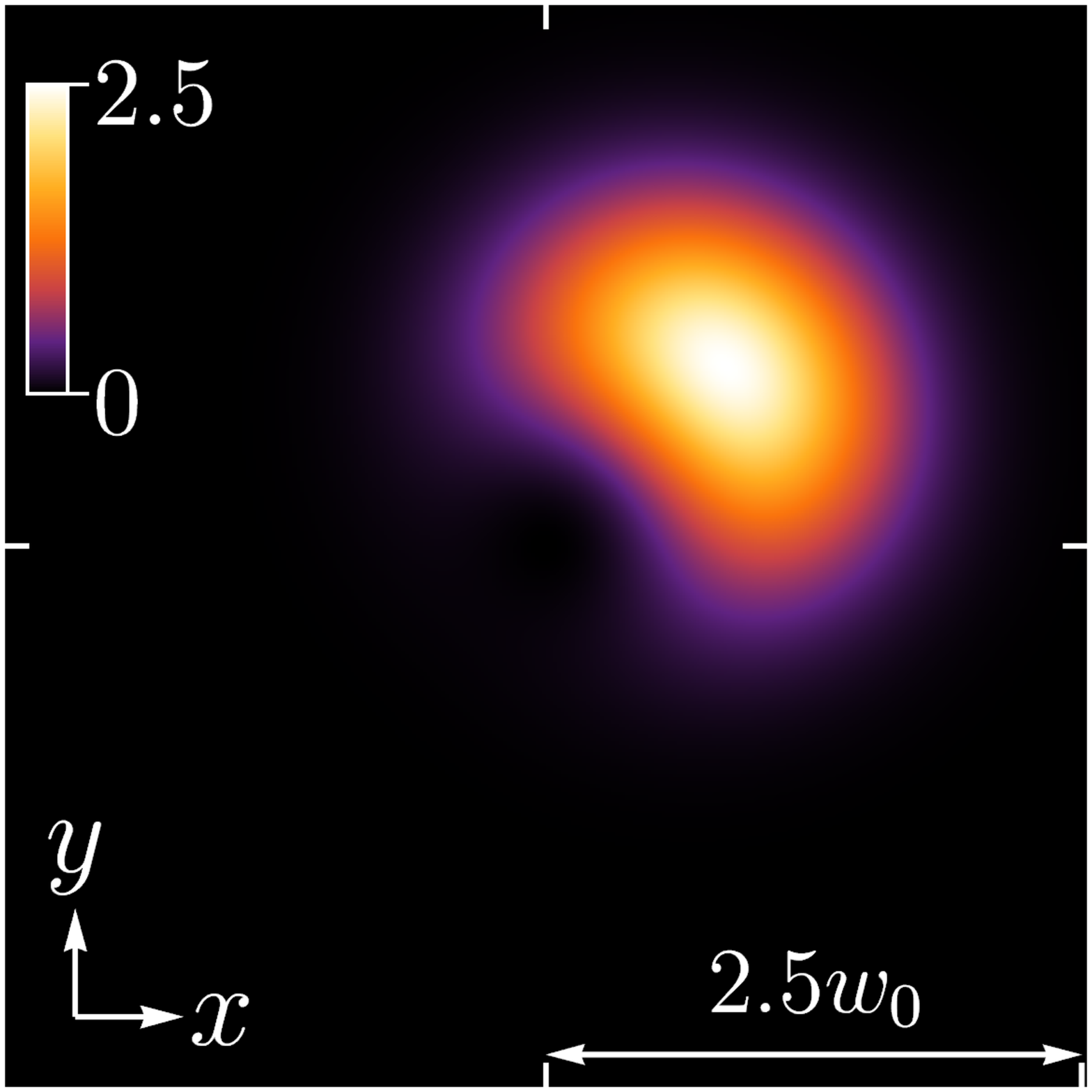}
		\label{fig:alg_z0_75_M45}}
	\caption{(Color online)  ($a$)-($d$) Transverse intensity pattern of asymmetric LG beam with topological charge $l=1$ for different values of asymmetry parameters $\delta$ and $ \beta $ at $ z=0 $ plane.}
	\label{fig:alg_z0}
\end{figure}

An LG beam profile propagating along the $z$-axis without any off-axis radial node and with positive helicity can be expressed in Cylindrical polar coordinate $(\rho,\varphi,z)$ \cite{allen_1992_lg} as
\begin{align}\label{eq:lgp0}
\mathcal{U}_{l}(\rho,\varphi,z)&=\frac{1}{\sqrt{l!}}\frac{w_0}{w(z)}\qty (\frac{\sqrt{2}\rho}{w(z)})^l 
e^{\im l \varphi} \exp(-\frac{\rho^2}{w^2(z)}) \times\nonumber\\
&\exp[\frac{\im k z \rho^2}{2(z^2+z_R^2)}-\im(l+1)\zeta(z)].
\end{align}
Here $w(z)=w_0 \sqrt{1+\dfrac{z^2}{z_R^2}} \text{ and } \zeta(z)=\tan[-1](\frac{z}{z_R})$, where $ w_0 $ is the waist of the beam, $ k $ is the wavenumber, and $ z_R = k w_0^2/2$ is called Rayleigh range. $l$ is the magnitude of topological charge or OAM quantum number per photon and it measures the amount of vorticity in the beam \cite{allen_1992_lg}.
The beam profile at the plane $ z=0 $ takes the form
\begin{align}
\mathcal{U}_l (\rho,\varphi)= \frac{1}{\sqrt{l!}}\qty(\frac{\sqrt{2}\rho}{w_0})^{l}
e^{\im l\varphi}\exp(-\frac{\rho^2}{w_0^2}).
\label{eq:lg0}
\end{align}
In order to include the asymmetric property in the beam, we consider complex valued shifts $\delta_x$ and $\delta_y$ in the transverse coordinate $x$ and $y$ respectively \cite{kovalev_2016_alg} such that \[x\rightarrow x - \delta_x \qq{and} y\rightarrow y-\delta_y.\] To retain the center of the vorticity at the axis of the beam, we assume  $\delta_x =-\im \delta_y = \delta e^{\im \beta}$, where $\delta$ and $ \beta $ are respectively the magnitude and the argument of $ \delta_x $. The shifted beam amplitude in the Cartesian coordinate takes the form as
\begin{align}
\mathcal{U}'_l(x,y)=\frac{\mathcal{A}}{ \sqrt{l!}}&\qty(\frac{\sqrt{2}}{w_0})^{l}\qty(x+\im y)^l\times \nonumber\\
& \exp\qty(-\frac{(x-\delta e^{\im \beta})^2 + (y-\im \delta e^{\im \beta})^2}{w_0^2}) ,
\label{eq:alg_xy}
\end{align}
where shift dependent parameter, $\mathcal{A}$ is
$$\mathcal{A}= \qty[\exp(\frac{2\delta^2}{w_0^2})L_l^0\qty(-\frac{2\delta^2}{w_0^2})]^{-1/2}$$ to keep the power of the beam profiles \eqref{eq:lg0} and \eqref{eq:alg_xy} unchanged.
Here $L_l^p$ is the associate Laguerre polynomial. However, transforming the Eq. \eqref{eq:alg_xy} into Cylindrical coordinate, we get
\begin{align}
\mathcal{U}'_l(\rho,\varphi)=& \frac{\mathcal{A}}{\sqrt{l!}}\qty(\frac{\sqrt{2}\rho}{w_0})^{l}
e^{\im l \varphi}\exp(-\frac{\rho^2}{w_0^2}+\frac{2 \rho \delta e^{\im \beta}}{w_0^2}e^{\im\varphi}).
\label{eq:alg_cy_z0}
\end{align}
Therefore, for nonzero values of $\delta$, the intensity profile of the beam does not preserve its symmetry in the Cylindrical coordinate, which is graphically illustrated in FIG. \ref{fig:alg_z0}.
In this figure, we have presented the variation of transverse intensity patterns of the beam at $z=0$ plane for different shift amplitudes $(\delta)$.  We have considered unit topological charge of the beam i.e., $l=1$, in all the plots of FIG. \ref{fig:alg_z0}. It can be clearly seen from the plots that the intensity distribution becomes more and more asymmetric in nature by increasing the magnitude of $ \delta $ and peak intensity position rotates with $ \beta$, but the center of the vortex position is still at the axis of the beam. FIG. \ref{fig:alg_z0_0_P0} shows the intensity pattern of regular LG beam, where we have taken the maximum intensity to be one. The  maximum values of the colorbars in the plots reflect that the asymmetricity increases the peak intensity of the beam and tries to confine the beam at a certain region in the transverse plane.

\begin{figure}[H]
	\centering
	\includegraphics[width=0.45\textwidth]{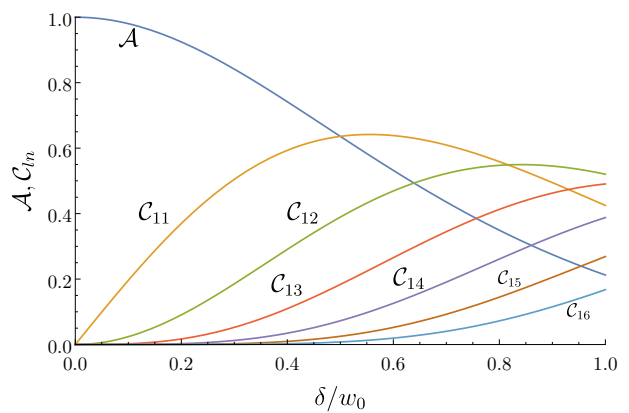}
	\caption{(Color online) Variation of the asymmetry coefficients $\mathcal{C}_{ln} $ and coefficient of primary component $\mathcal{A}$ with asymmetry parameter $ \delta $ for different value of mode index $ n $.}
	\label{fig:c_vari}
\end{figure}

In order to find the near field diffraction pattern, we have used Fresnel diffraction integral \cite{born_prin_optics} with the initial form $ \mathcal{U}'(\rho,\varphi,z=0) $ and we get
\begin{align}\label{eq:alg}
\mathcal{U}'_{l}(\rho,\varphi,z)&=\frac{\mathcal{A}}{\sqrt{l!}}\frac{w_0}{w(z)}\qty (\frac{\sqrt{2}\rho}{w(z)})^l 
e^{\im l \varphi}\exp(-\frac{s^2}{w^2(z)})\nonumber \\
&\times\exp(\frac{\im k z s^2}{2(z^2+z_R^2)} -\im(l+1)\zeta(z)),
\end{align}
where $ s^2= \rho^2 - 2\rho\delta e^{\im \beta} e^{\im \varphi} $. After simplifying (see Appendix \ref{ap:calc_step}), Eq. \eqref{eq:alg} takes the form as
\begin{align}\label{eq:alg_exp} 
	&\mathcal{U}'_{l}(\rho,\varphi,z)=\mathcal{A}~\mathcal{U}_l(\rho,\varphi,z)
	\exp[\frac{2\rho\delta e^{\im \beta}}{w_0 w(z)}e^{-\im \zeta(z)} e^{\im \varphi}]\nonumber\\
	&=\mathcal{A}~\mathcal{U}_l(\rho,\varphi,z)\qty [1+\sum_{n=1}^{\infty}\frac{1}{n!}
	\qty (\frac{2\rho\delta e^{\im \beta}}{w_0 w(z)})^n e^{-\im n \zeta(z)} e^{\im n\varphi}]\nonumber\\
	&=\mathcal{A}~\mathcal{U}_l(\rho,\varphi,z) +  \sum_{n=1}^{\infty} e^{\im n \beta} \mathcal{C}_{ln} 
	\mathcal{U}_{l+n}(\rho,\varphi,z),
\end{align}
where asymmetry coefficient $\mathcal{C}_{ln}= \frac{\mathcal{A}}{n!}\sqrt{\frac{(l+n)!}{l!}} \qty(\frac{\sqrt{2}\delta }{w_0})^n$. Therefore, by introducing the complex shift along $x$- and $y$-direction and keeping the center of the vortex along the axis of the beam, one can decompose the aLG beam as a superposition of an infinite number of coaxial LG beams of consecutive charges with different amplitudes. The exponential factor $ e^{\im n \beta} $ corresponds to a relative phase difference $ \beta $ between successive secondary LG beams. Here, we consider the value  of  $ \delta $  within the value of the beam waist. FIG. \ref{fig:c_vari} shows the variations of the asymmetry coefficients $\mathcal{C}_{ln}$ and the coefficient of primary component $(\mathcal{A})$. In this paper, we are considering the topological charge of the primary beam to be $l=1$ and the  maximum value of mode index, $n$, (in the sum of Eq. \eqref{eq:alg_exp}) to be 6 beyond which contributions from the secondary components are negligible with respect to the magnitude of complex shift in the unit of $w_0$. It is clearly seen from the figure that at the high shift amplitude of the beam, the asymmetry coefficients $(\mathcal{C}_{ln})$ dominant over the coefficient of primary component $(\mathcal{A})$.
\subsection{Interaction Hamiltonian}
We consider an aLG beam described above propagating along the $z$-axis of the laboratory frame interacting with a cold atom whose c.m. wavefunction has an extension 
comparable to the wavelength as well as the waist of the light beam. We also consider that the cold atomic system having the simplest form, composed of an electron of mass $m_e$ and charge $-e$; and 
a nucleus of mass $m_n$ and charge $+e$. 
The c.m. coordinate of the atomic system is $\vb{R}=\qty(m_e\vb{r}_e+m_n\vb{r}_n)/M$, 
with $M=m_e+m_n$ being the total mass where $\vb{r}_e$ and $\vb{r}_n$ being the position coordinate of the 
electron and nucleus respectively.
The atom experiences the local electric field as 
\begin{align}
\mathbf{E}(\bm{\rho},t)=\mathbf{E}_0~\mathcal{U}'_l (\rho,\varphi,z)e^{\im\qty (k z - \omega t)},
\label{eq:e_field}
\end{align} where $\omega$ is the angular frequency of the light beam and
$ \mathbf{E}_0 $ is the polarization vector.
At $t=0$, using Power Zineau Wooley (PZW) scheme,  the interaction Hamiltonian can be written  as \cite{babikar_2002_oam,mondal_2014_oam}
\begin{align}
H_I=e\frac{m_n}{M}\vb{r}\vdot\int_{0}^{1}\dd{\lambda} \mathbf{E}\qty(\vb{R}+\lambda\frac{m_n}{M}\vb{r}),
\label{eq:int_hamil}
\end{align}
 where the relative coordinate $\vb{r}=\vb{r}_e-\vb{r}_n$.
We assume the waist of the aLG beam in Eq. \eqref{eq:e_field} to be in the order of $\SI{e-5}{\meter}$, 
while the dimension of an electron orbital in an atom is of the order of a few
$\SI{}{\angstrom}$. 

Therefore, using the Taylor's expansion as $\abs{\vb{r}} \ll \abs{\vb{R}}$,
\begin{align}
\mathbf{E}\qty(\vb{R}+\lambda\frac{m_n}{M}\vb{r})=\vb{E}\qty(\vb{R})+
\lambda\frac{m_n}{M}\vb{r}\vdot\grad{\vb{E}\qty(\vb{R})}+\cdots \nonumber
\end{align}
Substituting this expression  in Eq. (\ref{eq:int_hamil}) and integrating, we get
\begin{align}
H_I= H_I^d+H_I^q+\cdots
\label{eq:int_hamil_2}
\end{align}
where $ H_I^d $ is the interaction Hamiltonian for the dipole transition and $ H_I^q $ is the same for the quadrupole transition which are given by
\begin{align}
H_I^d&=e\frac{m_n}{M}r\qty(\vu*{r}\vdot\vb{E}\qty(\vb{R}))\nonumber\\
&=\sqrt{\frac{4\pi}{3}}e \qty(\frac{m_n}{M})r\sum_{\sigma=0,\pm 1}\varepsilon_\sigma 
Y_1^\sigma(\theta,\phi)~
\mathcal{U}'_l(R_{\perp},\Phi, Z) e^{\im k Z}
\intertext{and}
H_I^q&=\frac{1}{2}e\qty(\frac{m_n}{M})^2
r^2\vu*{r}\vdot\qty(\vu*{r}\vdot\grad{\vb{E}\qty(\vb{R})})\nonumber\\
&=\frac{1}{2}\sqrt{\frac{4\pi}{3}}e \qty(\frac{m_n}{M})^2 r^2 \sum_{\sigma=0,\pm 1}\varepsilon_\sigma 
Y_1^\sigma(\theta,\phi)\nonumber\\
&\qquad\times\qty(\vu{r}\vdot\grad{\mathcal{U}'_l(R_{\perp},\Phi, Z)}) e^{\im k Z}
\end{align}
where we replaced $\vu*{r}\vdot\vb{E}_0$ by $\sum_{\sigma=0,\pm 1}\varepsilon_\sigma Y_1^\sigma(\theta,\phi)$
with 
$\varepsilon_{\pm1}=\mp \frac{1}{\sqrt{2}} (E_x \pm \im E_y) \text{ and }  \varepsilon_0=E_z $. Here $\sigma$ is the spin angular momentum (SAM) of the light. 
In the paraxial approximation, the $ E_z $ component is negligible. Here, $R_\perp$ is the projection of $\vb{R}$ 
on the transverse $XY$ plane.  

The population of atoms in the final condensate states depend on the transition matrix elements and they are derived using the form 
$\mathcal{M}_{i\rightarrow f} = \mel{\Upsilon_f}{H_I}{\Upsilon_i}$, where $\Upsilon$ 
denotes the unperturbed atomic states. 
We assume $\Upsilon\qty(\vb{R},\vb{r})=\Psi\qty(\vb{R})\psi\qty(\vb{r})$, where the c.m. wave 
function, $\Psi\qty(\vb{R})$,
depends on the external trapping potential; and the internal electronic wave function, 
$\psi\qty(\vb{r})$, can be considered
to be a highly correlated relativistic coupled-cluster orbital \cite{lindgren_1986_atomic,lindgren_1987_connectivity,dutta_2016_abinitio,das_2018_electron,biswas_2018_accurate}. 
The dipole matrix element is 
\begin{align}
\mathcal{M}^d_{i\rightarrow f} &= \sqrt{\frac{4\pi}{3}}e \qty(\frac{m_n}{M}) 
\sum_{\sigma=0,\pm 1}\varepsilon_\sigma \mel**{\psi_f}{r Y_1^\sigma(\theta,\phi)}{\psi_i} \nonumber\\
&
\times\qty[\mel**{\Psi_f}{\mathcal{A}~\mathcal{U}_l}{\Psi_i} + 
\mel**{\Psi_f}{\sum_{n=1}^{\infty} e^{\im n \beta}\mathcal{C}_{ln}~\mathcal{U}_{l+n}}{\Psi_i} ],
\label{eq:mat_ele_dip}
\end{align}
where $ \mathcal{U}_j=\mathcal{U}_j(R_{\perp},\Phi, Z)$, $j$ is any positive integer. The quadrupole matrix element is given by 
\begin{align}
	\mathcal{M}^q_{i\rightarrow f} &= \frac{1}{2} e \qty(\frac{m_n}{M})^2 \sum_{\sigma=0,\pm 1}\varepsilon_\sigma \Bigg[ \mel**{\psi_f}{r^2 Y_1^\sigma \sin\theta e^{\im \phi}}{\psi_i}\nonumber\\
	&\times\mel**{\Psi_f}{\qty( F_l+\sum_{n=1}^{\infty} e^{\im n \beta}\mathcal{C}_{ln} F_{l+n})e^{-\im \Phi}}{\Psi_i} \nonumber\\
	&-\mel**{\psi_f}{r^2 Y_1^\sigma \sin\theta e^{-\im \phi}}{\psi_i}\nonumber\\
	&\times\mel**{\Psi_f}{\qty (G_l+\sum_{n=1}^{\infty} e^{\im n \beta}\mathcal{C}_{ln} G_{l+n})e^{\im \Phi}}{\Psi_i}\Bigg].
	\label{eq:mat_ele_quad}
\end{align}
Here $ F$ and $G $ are function of $ \vb{R}$, defined as \[F_j = \mathcal{U}_j(R_\perp,\Phi,Z) R_\perp^{-1}\qty (j-\frac{R_\perp^2}{w_0 w(Z)}e^{-\im \zeta(Z)})\] and  \[G_j = \frac{\mathcal{U}_j(R_\perp,\Phi,Z) R_\perp}{w_0 w(Z)}e^{-\im \zeta(Z)}.\]

Here, we consider the atomic system in a constant $z$-plane, therefore we omitted the gradient over $ z $ in $ \grad{\mathcal{U}'}$. Let us discuss each of the terms in Eqs. \eqref{eq:mat_ele_dip} and \eqref{eq:mat_ele_quad} to understand the mechanism of transferring the OAM ($l$) and SAM ($\sigma$) from aLG beam to the cold atom. The terms which appear in the summation with index $n$ of the Eqs. \eqref{eq:mat_ele_dip} and \eqref{eq:mat_ele_quad}, signify the  effects of the asymmetric nature of the aLG beam on the interacting atom. In the dipole transition, the first term in the square bracket denotes the usual interaction of the LG beam with the atom, where the OAM of the beam transfers to the c.m. of the atom which is already shown in the literatures \cite{romero_2002_quantum,jauregui_2004_rota,mondal_2014_oam,bhowmik_2016_interaction,bhowmik_2018_density}. In addition to the first term, the second term shows the transfer of multiple vorticities, $l+n$ (where $n=1,2,3,\cdots,\infty $) to the c.m. of the atom due to the asymmetric property of the beam. However, in all the transition channels in Eq. \eqref{eq:mat_ele_dip}, the SAM of the beam always goes to the electronic motion of the atom and satisfies the selection rule of the transition. 

In case of the quadrupole transition, as suggested by Mondal \emph{et al.} \cite{mondal_2014_oam}, one unit of OAM from the beam is possible to transfer to the electronic motion  via the c.m. of the atom and modifies the selection rule of transition. Therefore, one can see from the Eq. \eqref{eq:mat_ele_quad} that two excited electronic states are coupled to the initial electronic state through the quadrupole transition. These coupling of two excited states happens by transferring $\sigma+1$ and $\sigma-1$ unit of angular momenta to the electronic motion of the atom unlike in the case of dipole transitions where the only spin component, $\sigma$ unit of angular momentum, goes to the electronic motion. In the former case, $l-1,l,l+1,\cdots$ unit of multiple vorticities will be transferred to the c.m. of the atom but in the latter case $l+1, l+2, l+3,\cdots$ unit of multiple vorticities will go the c.m. of the atom. These multiple vorticities arise due to the asymmetric nature of the aLG beam, having different amplitudes depending on the asymmetric coefficients $\mathcal{C}_{ln}$. In the next subsection, we will discuss how this theoretical model of interaction of single aLG beam with cold atom can be employed to create the superposition of vortex states in BEC.

\subsection{Creation of superposition of vortex states in BEC}
For a disk-shaped condensate, the three-dimensional Gross-Pitaevskii (GP) equation 
can be reduced to a two-dimensional GP equation by assuming that the time evolution 
does not cause any excitation along the $z$-direction. Experimentally, one can achieve this disk-shaped condensate by applying a very strong trapping potential along the transverse direction compared to the $x$- and $y$-direction, \emph{i.e.} $\hbar \omega_z >> \hbar \omega_\perp$ \cite{bao_2003_numerical}.
We have considered that the BEC is trapped in a 2-D harmonic potential where the initial and final stationary states of the c.m. motion of atoms can be written as 
$$
\Psi_i(\vb{R})=\Psi_i(R_{\perp})e^{\im \kappa_i \Phi} \qq{,} 
\Psi_f(\vb{R})=\Psi_f(R_{\perp})e^{\im \kappa_f \Phi}
$$
where $i (f)$ stands for initial (final) and $\kappa $ is the quantum circulation of atoms about the $ z $ axis.   $ \kappa\ne0 $ represents 
vortex states of the BEC.
\begin{figure}[H]
	\centering
	\subfloat{\includegraphics[width=0.45\textwidth]{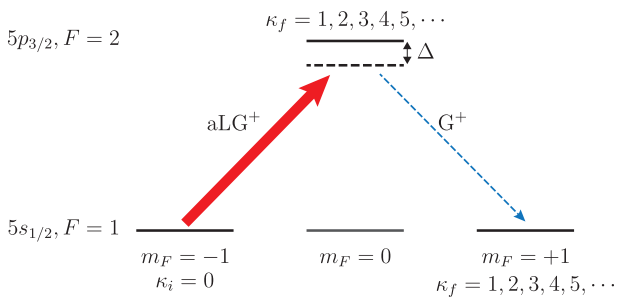}}
	\caption{(Color online) Energy-level scheme of the two-photon dipole transitions. The atomic states show the $ \isotope[87]{Rb} $ hyperfine states. Atoms are initially trapped in $\ket{5s_\frac{1}{2},F=1,m_F=-1}$. $ \Delta $ represents two-photon detuning. The superscript `+' on $\mathrm{aLG}^+$ and $ \mathrm{G}^+ $ represents polarization $ \sigma = +1 $. Here, OAM of the $\mathrm{aLG}^+$ beam is $l=1$.}
	\label{fig:dip_tran}
\end{figure}
We have considered a left circularly $(\sigma=+1 )$ polarized $\mathrm{aLG}^+$ beam with wavelength 
$\SI{780.03}{\nano\metre}$ and OAM$=l$ unit interacting with the atoms at BEC  prepared in the ground state $\ket{5s_\frac{1}{2},F=1,m_F=-1}$. The $\mathrm{aLG}^+$ pulse induces dipole transitions in atoms as given in Eq. \eqref{eq:mat_ele_dip}. The electronic portions $\mel{\psi_f(\vb{r})}{r Y_1^{+1}}{\psi_i(\vb{r})}$  on the right hand side of Eq. \eqref{eq:mat_ele_dip} indicate that the intermediate electronic state (as shown in the FIG. \ref{fig:dip_tran}), $\psi_f$ will be $\ket{5p_\frac{3}{2},F=2,m_F=0}$. In order to bring back the atoms
in the  different hyperfine sublevel $ \qty(\ket{5s_\frac{1}{2},F=1,m_F=+1})$ of the ground state using two-photon stimulated Raman transition, we have used a
Gaussian pulse which is co-propagating with the $\mathrm{aLG}^+$ beam having appropriate frequency and polarization. Assuming the initial vorticity of the BEC, $\kappa_i=0$ and employing Eq. \eqref{eq:mat_ele_dip}, one can create a superposition of vortex states at the final hyperfine sublevel, $ \ket{5s_\frac{1}{2},F=1,m_F=+1} $ with the consecutive vorticities $\kappa_f = l, l+1, l+2, l+3, l+4,\cdots$. This superposition of the vortex states can be expressed as \cite{bhowmik_2016_interaction}
\begin{align}
\label{eq:alpha}
\Psi(R_{\perp},\Phi,t) &= e^{-\im \frac{\mu}{\hbar} t} \sum_{j=l}^{\infty} \alpha_{j} f_{j}(R_\perp)  e^{\im j \Phi},
\end{align}
where $ \mu $ is the chemical potential of the system. $ f_{j}(R_\perp) $ is the radial function of the final wavefunction with vorticity $ j $. $ \alpha_{i}$s' are constants depending on the coefficient $\mathcal{C}_{ln} $ and two-photon Rabi frequencies of corresponding vortex states, with  
$ \sum_{j=l}^{\infty}  \abs{\alpha_{j}}^2 = 1.$

\begin{figure}[H]
	\centering
	\includegraphics[width=0.45\textwidth]{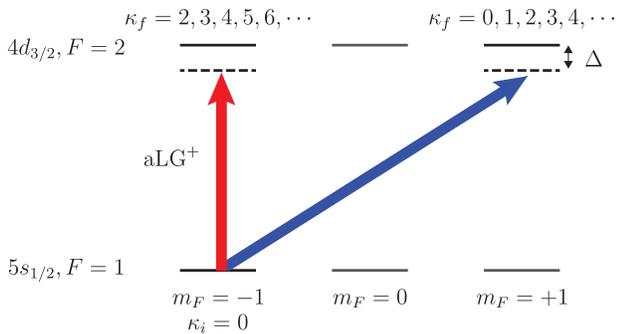}
	\caption{(Color online) Energy-level scheme of single-photon quadrupole transition with topological charge, $l=1$ of $\mathrm{aLG}^+$ beam.}
	\label{fig:quad_tran}
\end{figure}

For electric quadrupole transition, we assume an $\mathrm{aLG}^+$ beam with wavelength \SI{516.51}{\nano\meter} with left circular polarization  interacting with the BEC trapped in the state, $\ket{5s_\frac{1}{2},F=1,m_F=-1}$. As derived in Eq. \eqref{eq:mat_ele_quad} and depicted in FIG. \ref{fig:quad_tran}, $\pm 1$ unit of OAM is transferred to the electronic motion of the atom resulting  two different types of quadrupole transitions with the changes at $m_F$, $\Delta m_F=+2$ and $0$. In the quadrupole transitions, presented in Eq. \eqref{eq:mat_ele_quad}, there are two electronic transition parts. They are the matrix elements $\mel{\psi_f(\vb{r})}{r Y_1^{+1}e^{\im \phi}}{\psi_i(\vb{r})}$ and $\mel{\psi_f(\vb{r})}{r Y_1^{+1}e^{-\im \phi}}{\psi_i(\vb{r})}$ highlighted with blue and red arrows in the FIG. \ref{fig:quad_tran}. According to these matrix elements, the final states will be $ \ket{4d_{3/2},F=2,m_F=+1}$ with multiple vorticity $ \kappa_f=l-1,l,l+1,l+2,l+3,\cdots$ and $ \ket{4d_{3/2},F=2,m_F=-1}$ with $ \kappa_f = l+1,l+2,l+3,l+4,l+5,\cdots$, respectively. As the transition probability of the quadrupole transition is always very less compare to the  dipole transition, we have discussed only  the  single-photon transitions for quadrupole case in the rest of the paper. However, one can create here a superposition of vortex states using a suitable choice of  Gaussian beam through the two-photon Raman transitions.
\subsection{First order spatial correlation of the condensate}

The equal-time correlation of a field at zero-temperature can be expressed as \cite{naraschewski_1999_spatial}
\begin{align}
G^1(\vb{R}_1,\vb{R}_2)= \expval{\Psi^{\dagger}(\vb{R}_1)\Psi(\vb{R}_2)}
\label{eq:corr1}
\end{align}
In the point of view of many experiments, the first order degree of coherence at two dimensional separation $\vb{r}_c = \vb{R}_1 - \vb{R}_2$  is calculated  in the form of surface integral (integration has been carried out over the spatial region $ \vb{R}_c = \frac{\vb{R}_1 + \vb{R}_2}{2}$) at $z=0$ as 

\begin{align}
& g^{(1)}(\vb{r}_c) = \nonumber \\
& \frac{\mathlarger{\int} \dd{\vb{R}_c}G^1(\vb{R}_c-\frac{\vb{r}_c}{2},\vb{R}_c+\frac{\vb{r}_c}{2})}{\mathlarger{\int} \dd{\vb{R}_c}\sqrt{G^1(\vb{R}_c-\frac{\vb{r}_c}{2},\vb{R}_c-\frac{\vb{r}_c}{2})}\sqrt{G^1(\vb{R}_c+\frac{\vb{r}_c}{2},\vb{R}_c+\frac{\vb{r}_c}{2})}} \nonumber \\
&=\frac{\mathlarger{\int} \dd{\vb{R}_c}\Psi^*(\vb{R}_c-\frac{\vb{r}_c}{2}) \Psi(\vb{R}_c+\frac{\vb{r}_c}{2})}{\mathlarger{\int} \dd{\vb{R}_c}\abs{\Psi(\vb{R}_c-\frac{\vb{r}_c}{2})}\abs{\Psi(\vb{R}_c+\frac{\vb{r}_c}{2})}} .
\label{eq:coher}
\end{align}

\section{Numerical Results and Discussion}\label{sec:res_dis}
 We consider that an aLG beam (with $l=1$) interacts with a non-rotating ($ \kappa_i=0 $) 2D BEC  having $ 10^5 $ number of $\isotope[87]{Rb}$ atoms, trapped in the ground state $\qty(\ket{5s_\frac{1}{2},F=1,m_F=-1})$. The Rabi frequencies of dipole and quadrupole transitions using Eqs. \eqref{eq:mat_ele_dip} and \eqref{eq:mat_ele_quad} are now evaluated numerically by solving the c.m. and electronic wavefunctions obtained from 2D GP equation \cite{dalfovo_1996_bosons,bao_2003_numerical}  and relativistic coupled-cluster theory \cite{lindgren_1986_atomic,das_2018_electron}, respectively. For a disk-shaped condensate, axial trap frequency $ \omega_z $ is very large compared to the cylindrically
symmetric radial frequency $ \omega_\perp $ \cite{bao_2003_numerical}. The asymmetry parameter of the harmonic 
trap is $ \gamma = \frac{\omega_z}{\omega_\perp}\gg 1 $. 
For $ \omega_\perp /{2 \pi} = \SI{20}{Hz}$ the corresponding characteristic length is 
$a_\perp = \SI{2.4114}{\micro\meter}$. For maximum interaction region of beam with trapped BEC, the waist $ w_0 $ of the beam is set to be nearly five times of $ a_\perp $ which is $\SI{1.2e-5}{m} $ and the 
intensity $ I = \SI{e2}{\watt\cm^{-2}} $. The amplitude of the aLG beam $ \varepsilon $ is related to the
intensity $I$ by the relation $I=\frac{\epsilon_0 c}{2}\varepsilon^2$, where $\epsilon_0$ is the free space 
permittivity. To calculate the two-photon Rabi frequencies for the dipole transition, we consider
that co-propagating aLG and a Gaussian beam with left circularly polarization ($\sigma=+1 $) are incident on the trapped BEC.  As shown in FIG. \ref{fig:dip_tran}, the atoms which will participate in the two-photon transition will reach  the final electronic state, $ \ket{5s_\frac{1}{2},F=1,m_F=+1}$, of the condensate. Here Gaussian beam is detuned from D2 line by $\Delta = \SI{-1.5}{\giga\Hz}$ which is enough to prevent the destructive incoherent heating of the condensate due to spontaneous decay of excited states. Anyway, applying the two-photon transition using aLG and Gaussian beams, the superposition of vortex states are produced at another hyperfine level $ \ket{5s_\frac{1}{2},F=1,m_F=+1}$ of ground state. According to the Eq. \eqref{eq:mat_ele_dip}, the created multiple vorticities  have the quantum circulations,  $\kappa_f= 1,2,3,4,5,\cdots$. Apart from the  $\kappa_f=1$, others are generated due to the asymmetric property of the aLG beam. As the magnitudes of two-photon Rabi frequencies which have quantum circulation $\kappa_f>5$ are very small, they are neglected in our calculations and demonstration.

\subsection{Superposition of matter-vortex states through two-photon dipole transitions}

\begin{table}[H]
	\renewcommand{\arraystretch}{1.3}
	\renewcommand\tabcolsep{0.5em}
	\centering
	\caption{(Color online) Magnitudes of two-photon Rabi frequencies (in $ \SI{}{\Hz} $) in the dipole interaction for different shift parameter, $ \delta/w_0$}
\centering
{\scriptsize
		\begin{tabular}{cSSSSS}
			\hline
			\multirow{2}{*}{$\delta/w_0$} & \multicolumn{5}{c}{Two-photon Rabi freq. $\Omega_d$ for final vortex 
				states}\\
			&$\kappa_f = 1 $&$\kappa_f = 2$&$\kappa_f = 3$&$\kappa_f = 4$&$\kappa_f = 5$\\
			\hline \hline

			0     & 3.97E+09 & 0.00E+00 & 0.00E+00 & 0.00E+00 & 0.00E+00 \\
			0.1   & 3.89E+09 & 6.32E+08 & 5.82E+07 & 3.90E+06 & 2.09E+05 \\
			0.2   & 3.67E+09 & 1.19E+09 & 2.20E+08 & 2.94E+07 & 3.16E+06 \\
			0.3   & 3.34E+09 & 1.63E+09 & 4.50E+08 & 9.04E+07 & 1.46E+07 \\
			0.4   & 2.94E+09 & 1.91E+09 & 7.05E+08 & 1.89E+08 & 4.06E+07 \\
			0.5   & 2.52E+09 & 2.05E+09 & 9.44E+08 & 3.16E+08 & 8.49E+07 \\
			0.6   & 2.11E+09 & 2.06E+09 & 1.14E+09 & 4.58E+08 & 1.47E+08 \\
			0.7   & 1.73E+09 & 1.96E+09 & 1.27E+09 & 5.95E+08 & 2.23E+08 \\
			0.8   & 1.39E+09 & 1.80E+09 & 1.33E+09 & 7.12E+08 & 3.05E+08 \\
			0.9   & 1.09E+09 & 1.59E+09 & 1.32E+09 & 7.98E+08 & 3.85E+08 \\
			1     & 8.43E+08 & 1.37E+09 & 1.26E+09 & 8.46E+08 & 4.54E+08 \\

			\hline \hline
		\end{tabular}%
	}
	\label{tab:dip_rabi}%
\end{table}%

\begin{figure}[h]
	\centering
	\includegraphics[width=0.47\textwidth]{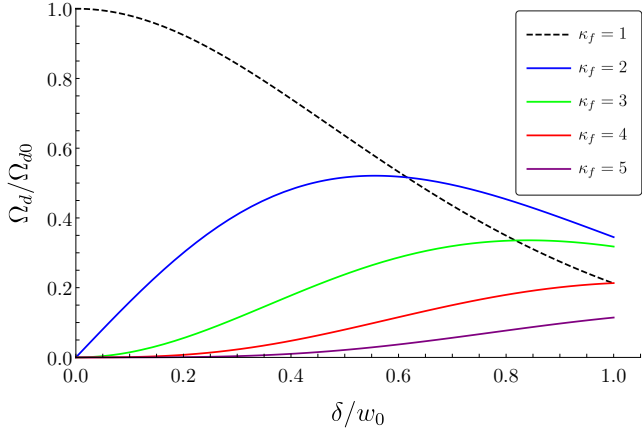}
	\caption{(Color online) Variation of two-photon Rabi frequency $ \Omega_d $ with asymmetry ($\delta \ne 0$)  relative to the Rabi frequency  $ \Omega_{d0} $ without asymmetry ($\delta = 0$) for different $ \delta/w_0 $ using dipole approximation.$w_0$ is the beam waist and $\Omega_{d0} = \SI{3.97e9}{\Hz}$. $\kappa_f$ denotes (see legends) different vortex states. }
	\label{fig:dip_rabi}
\end{figure}

Table \ref{tab:dip_rabi} presents the two-photon Rabi frequencies for the dipole transitions $(\Omega_d)$ for the quantum circulations of the atoms, $\kappa_f=1$ to $5$.  Also, FIG. \ref{fig:dip_rabi} displays the variation of the ratio $\Omega_d/\Omega_{d0}$ with the  shift parameter $\delta/\omega_0$ of the beam, where $\Omega_{d0}$ is the magnitude of the Rabi frequency of primary transition in absence of  the shift  ($\delta=0$).  It is clear from the table as well as from the figure that the Rabi frequency of the primary transition corresponding to $\kappa_f=1$ decreases monotonically with the increase of the shift parameter. The Rabi frequencies for the secondary transitions with  higher vorticities of final states  increases initially with the shift and reach to the highest value of $\Omega_d$ at different values of the shift parameters due to the asymmetric nature of the beam. At those peak positions, the Rabi frequencies of the secondary transitions slightly differ from the Rabi frequency of primary transition with $\kappa_f=1$.  It signifies an interesting feature that, for the high shift parameters of the beam, the number of atoms interacting with the additional vorticities ($n\ge 1$ in Eq. \eqref{eq:mat_ele_dip})  of the beam can be comparative to primary vorticity.

\begin{table}[h]
	\renewcommand{\arraystretch}{1.3}
	\renewcommand\tabcolsep{0.5em}
	\centering
	\caption{Magnitude of single-photon Rabi frequencies (in $ \SI{}{\Hz} $) in  the quadrupole interaction for $ \Delta m_F =+2 $ with different shift parameter, $ \delta/w_0$}
	\centering
	{\scriptsize
		\begin{tabular}{cSSSSS}
			\hline
			\multirow{2}{*}{$\delta/w_0$} & \multicolumn{5}{c}{Single-photon Rabi freq. $\Omega_q$ for final vortex 
				states}\\
			&$\kappa_f = 0 $&$\kappa_f = 1$&$\kappa_f = 2$&$\kappa_f = 3$&$\kappa_f = 4$\\
			\hline \hline
			
				0     & 2.59E+04 & 0.00E+00 & 0.00E+00 & 0.00E+00 & 0.00E+00 \\
				0.1   & 2.54E+04 & 9.03E+03 & 1.21E+03 & 1.05E+02 & 6.86E+00 \\
				0.2   & 2.39E+04 & 1.70E+04 & 4.58E+03 & 7.93E+02 & 1.04E+02 \\
				0.3   & 2.18E+04 & 2.33E+04 & 9.38E+03 & 2.44E+03 & 4.77E+02 \\
				0.4   & 1.92E+04 & 2.73E+04 & 1.47E+04 & 5.09E+03 & 1.33E+03 \\
				0.5   & 1.65E+04 & 2.93E+04 & 1.97E+04 & 8.52E+03 & 2.78E+03 \\
				0.6   & 1.38E+04 & 2.94E+04 & 2.37E+04 & 1.23E+04 & 4.82E+03 \\
				0.7   & 1.13E+04 & 2.81E+04 & 2.64E+04 & 1.60E+04 & 7.31E+03 \\
				0.8   & 9.04E+03 & 2.57E+04 & 2.77E+04 & 1.92E+04 & 1.00E+04 \\
				0.9   & 7.11E+03 & 2.28E+04 & 2.76E+04 & 2.15E+04 & 1.26E+04 \\
				1     & 5.50E+03 & 1.96E+04 & 2.63E+04 & 2.28E+04 & 1.49E+04 \\
			
			\hline \hline
		\end{tabular}%
	}
	\label{tab:quad_Rabi1}%
\end{table}%
\begin{figure}[h]
	\centering
	\includegraphics[width=0.47\textwidth]{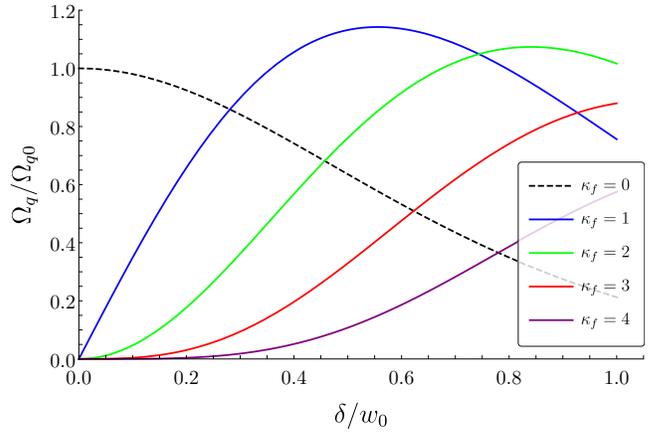}
	\caption{(Color online) Variation of the ratio of quadrupole Rabi frequency $  \Omega_q$ with asymmetry ($\delta \ne 0)$  to the Rabi frequency $\Omega_{q0}$ without asymmetry ($\delta = 0$ ) for different vortex states $\kappa_f$(see legends) with $ \delta/w_0 $ for $ \Delta m_F=+2~(\Omega_{q0}=\SI{2.59e4}{\Hz}).$ $w_0$ represents the beam waist.}
	\label{fig:quad_rabi_1}
\end{figure}

\subsection{Dependance of quadrupole Rabi frequencies on the shift parameter}

\begin{figure}[H]
	\centering
	\includegraphics[width=0.47\textwidth]{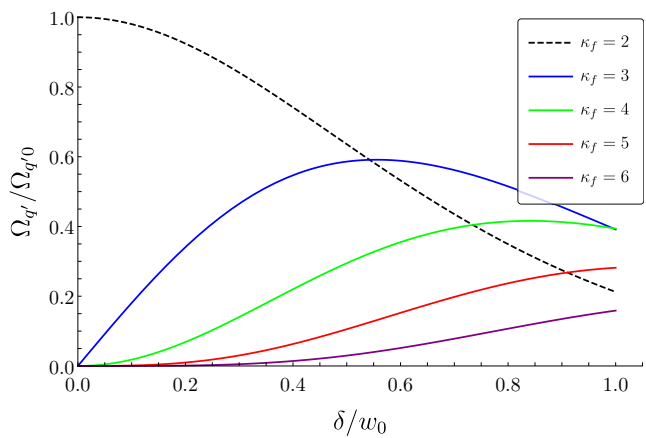}
	\caption{(Color online) Variation of ratio $ \Omega_{q'}/\Omega_{q'0}$ quadrupole Rabi frequency for different vortex states with $ \delta/w_0 $ for $ \Delta m_F=0~(\Omega_{q'0}=\SI{2.58e4}{\Hz}). $ $\kappa_f$ denotes the vorticity of the final state.}
	\label{fig:quad_rabi_2}
\end{figure}

Table \ref{tab:quad_Rabi1} and \ref{tab:quad_Rabi2} present the variation of Rabi frequencies with the shift parameter for the single-photon quadrupole transitions corresponding to selection rule $\Delta m_F=+2$ and 0, respectively. FIG. \ref{fig:quad_rabi_1} and \ref{fig:quad_rabi_2} correspond to Table \ref{tab:quad_Rabi1} and \ref{tab:quad_Rabi2}, respectively.   
Due to the transfer of one unit OAM to the electronic motion, the primary quadrupole transitions generate electronic states $ \ket{4d_{3/2},F=2,m_F=+1}$ and $ \ket{4d_{3/2},F=2,m_F=-1}$ of the condensate (see FIG. \ref{fig:quad_tran}) with the vorticities  $\kappa_f=0$ and $2$, respectively. The additional secondary vorticities are  produced from the asymmetric property of the beam. As shown in the figures,  Rabi frequencies of the primary transitions for both the cases  monotonically decrease with the shift parameter, similar to the case of dipole transition discussed above. For the quadrupole transition with $\Delta m_F=0$, the Rabi frequency of the primary transition (with $\kappa_f=2$) can be smaller than the Rabi frequencies of the next three secondary transition ( with $\kappa_f=3, 4$ and $5$) for high shift parameters of the beam. However, the situation is different for the quadrupole transition satisfying the selection rule $\Delta m_F=2$. In this case, at high shift parameter of the beam, the Rabi frequency of the primary quadrupole transition (with $\kappa_f=0$) is significantly smaller in comparison to the all four secondary transitions presented in the FIG. \ref{fig:quad_rabi_1}.  The interesting feature in FIG. \ref{fig:quad_rabi_1} is that at the shift parameters close to 0.6 and 0.8,  the quadrupole Rabi frequencies for the secondary transitions with $\kappa_f=1$ and $2$ can have appreciable large values compared to the primary transition ($\Omega _{q0}$). These enhancements of the amplitudes of  the quadrupole Rabi frequencies arise due to term $ (j-\frac{R^2_\perp}{w_0^2}) $ in the function $F_j$ in Eq. \eqref{eq:mat_ele_quad} and such effects have not been observed to date.
\begin{table}[H]
	\renewcommand{\arraystretch}{1.3}
	\renewcommand\tabcolsep{0.5em}
	\centering
	\caption{Magnitudes of single-photon Rabi frequencies (in $ \SI{}{\Hz} $) in the quadrupole transition for $ \Delta m_F =0 $ with different shift parameter, $ \delta/w_0$}
	\centering
	{\scriptsize
		\begin{tabular}{cSSSSS}
			\hline
			\multirow{2}{*}{$\delta/w_0$} & \multicolumn{5}{c}{Single-photon Rabi freq. $\Omega_{q'}$ for final vortex states}\\
			&$\kappa_f = 2 $&$\kappa_f = 3$&$\kappa_f = 4$&$\kappa_f = 5$&$\kappa_f = 6$\\
			\hline \hline
			
				0     & 2.58E+04 & 0.00E+00 & 0.00E+00 & 0.00E+00 & 0.00E+00 \\
				0.1   & 2.53E+04 & 4.66E+03 & 4.69E+02 & 3.35E+01 & 1.89E+00 \\
				0.2   & 2.39E+04 & 8.79E+03 & 1.77E+03 & 2.53E+02 & 2.85E+01 \\
				0.3   & 2.17E+04 & 1.20E+04 & 3.62E+03 & 7.77E+02 & 1.31E+02 \\
				0.4   & 1.91E+04 & 1.41E+04 & 5.68E+03 & 1.62E+03 & 3.66E+02 \\
				0.5   & 1.64E+04 & 1.51E+04 & 7.60E+03 & 2.72E+03 & 7.66E+02 \\
				0.6   & 1.37E+04 & 1.52E+04 & 9.16E+03 & 3.93E+03 & 1.33E+03 \\
				0.7   & 1.12E+04 & 1.45E+04 & 1.02E+04 & 5.11E+03 & 2.02E+03 \\
				0.8   & 9.01E+03 & 1.33E+04 & 1.07E+04 & 6.12E+03 & 2.76E+03 \\
				0.9   & 7.09E+03 & 1.18E+04 & 1.06E+04 & 6.85E+03 & 3.48E+03 \\
				1     & 5.48E+03 & 1.01E+04 & 1.02E+04 & 7.27E+03 & 4.09E+03 \\
			
			\hline \hline
		\end{tabular}%
	}
	\label{tab:quad_Rabi2}%
\end{table}%

\begin{figure}[h!]
	\centering
	\subfloat[$\delta/w_0 = 0$]{\includegraphics[width=0.24\textwidth]{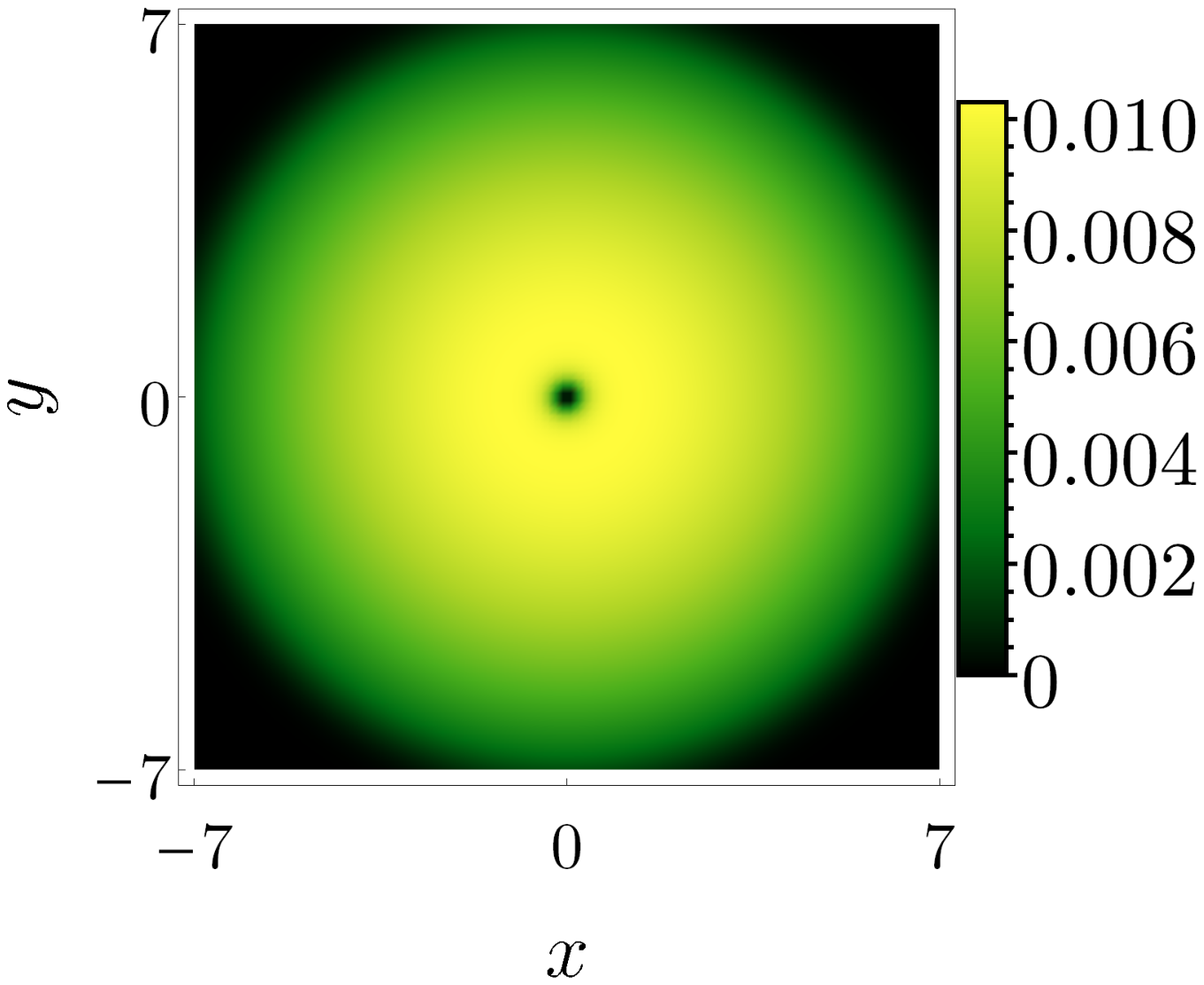}}
	\subfloat[$\delta/w_0 = 0.3$]{\includegraphics[width=0.24\textwidth]{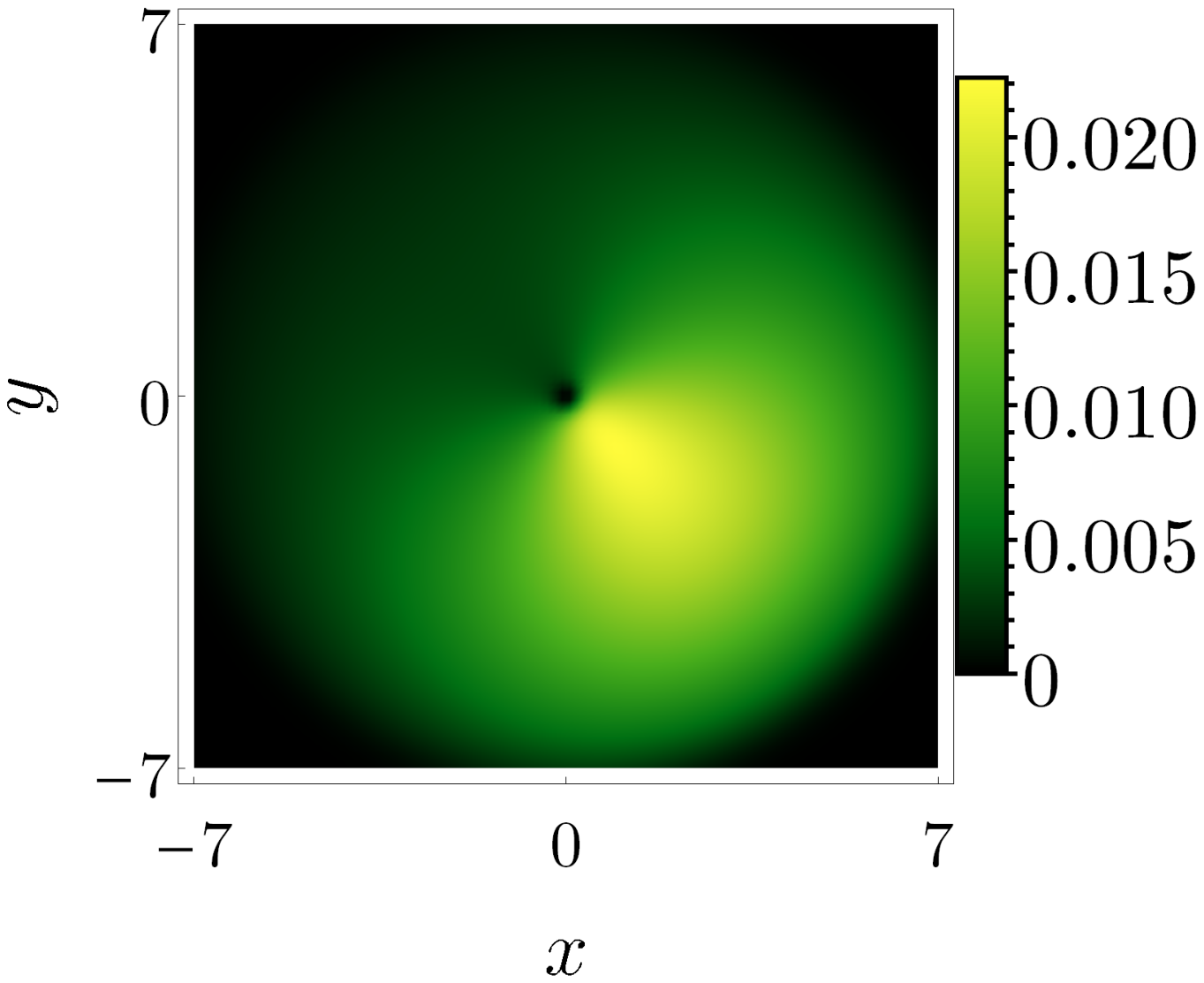}}
	\vspace{0.2cm}
	\subfloat[$\delta/w_0 = 0.6$]{\includegraphics[width=0.24\textwidth]{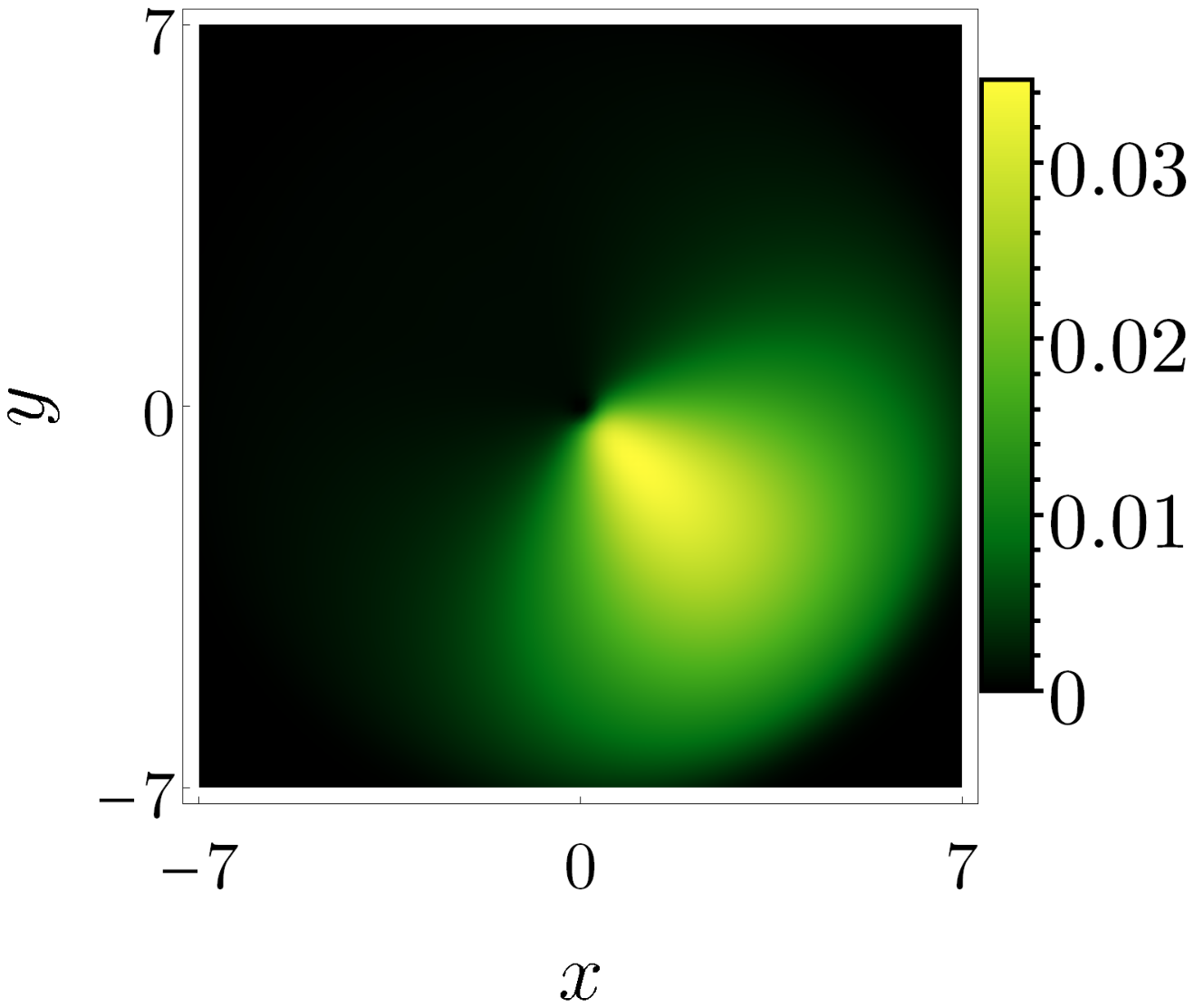}}
	\subfloat[$\delta/w_0 = 0.9$]{\includegraphics[width=0.24\textwidth]{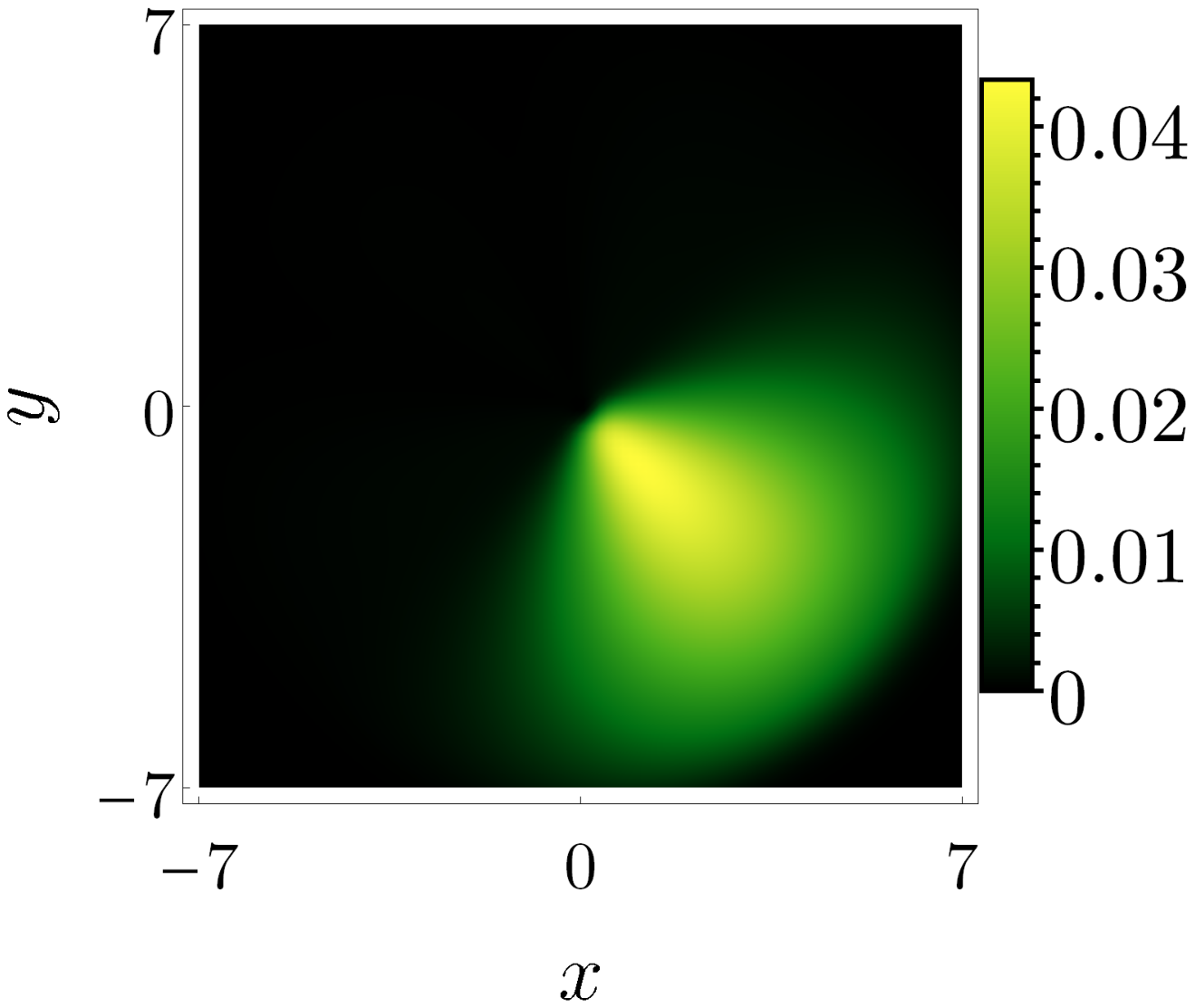}}
	\caption{(Color online) ($a$)-($d$) Particle density in BEC due to the superposition of vortex states with $ \beta = \pi/4 $ for different value of asymmetry parameters $ \delta $. $w_0$ represents the beam waist of the beam and $ x $ and $ y $ are the transverse coordinates of the trap in the unit of $ a_\perp $.}
	\label{fig:bec_vortex_super}
\end{figure}

As shown in the FIG. \ref{fig:dip_tran},  the condensate ground state with atoms having electronic state $\ket{5s_\frac{1}{2},F=1,m_F=+1}$ contains multiple vorticity. Therefore, a superposition of matter-wave vortex states is generated from  single aLG beam in the non-vortex BEC. The pattern of the superposition depends on the populations of each of the vortex states which are calculated from the probability amplitudes $\alpha_1, \alpha_2,\alpha_3, \alpha_4 \text{ and } \alpha_5$ corresponding to the macroscopic matter vortices,  $\kappa_f$ =1, 2, 3, 4 and 5, respectively, (see Eq. \eqref{eq:alpha}). These probability amplitudes are determined from the two-photon Rabi frequencies of the dipole transition and each of the $\alpha$'s carry a phase term $e^{in\beta}$. The variations of the $\alpha$'s with the shift parameters have been shown in Table \ref{tab:alpha}. Using the Eq. \eqref{eq:alpha} and Table \ref{tab:alpha}, we have estimated the superposition of matter-wave vortex states for different shift parameters and presented in FIG. \ref{fig:bec_vortex_super}. The patterns in this figure indicate that the matter density gets concentrated at a certain region of space with the increasing values of shift parameters. Controlling the parameters $\delta$ and $\beta$, we can manipulate the position of peak particle density in BEC. 

\begin{table}[H]
	\renewcommand{\arraystretch}{1.3}
	\renewcommand\tabcolsep{0.7em}
	\centering
	\caption{Magnitude of the probability amplitudes $ \alpha $ for different $ \delta $ at $z=0$}
	\centering
	{\scriptsize
	\begin{tabular}{cSSSSS}
		\hline
		$ \delta/w_0 $	 & $ \alpha_{1} $ & $ \alpha_{2} $ & $ \alpha_{3} $ & $ \alpha_{4} $ & $ \alpha_{5} $\\
		\hline \hline
    0     & 1     & 0     & 0     & 0     & 0 \\
	0.1   & 0.9870 & 0.1603 & 0.0148 & 0.0010 & 0.0001 \\
	0.2   & 0.9495 & 0.3084 & 0.0568 & 0.0076 & 0.0008 \\
	0.3   & 0.8922 & 0.4347 & 0.1202 & 0.0242 & 0.0039 \\
	0.4   & 0.8210 & 0.5333 & 0.1966 & 0.0527 & 0.0113 \\
	0.5   & 0.7420 & 0.6025 & 0.2776 & 0.0931 & 0.0250 \\
	0.6   & 0.6606 & 0.6437 & 0.3559 & 0.1432 & 0.0461 \\
	0.7   & 0.5807 & 0.6601 & 0.4259 & 0.1999 & 0.0751 \\
	0.8   & 0.5052 & 0.6563 & 0.4839 & 0.2596 & 0.1114 \\
	0.9   & 0.4358 & 0.6369 & 0.5283 & 0.3188 & 0.1539 \\
	1     & 0.3733 & 0.6063 & 0.5588 & 0.3747 & 0.2010 \\
		\hline \hline
	\end{tabular}%
	}
	\label{tab:alpha}%
\end{table}

\subsection{Spatial degree of coherence of the superposed condensate }
\begin{figure}[H]
	\centering
	\includegraphics[width=0.5\textwidth]{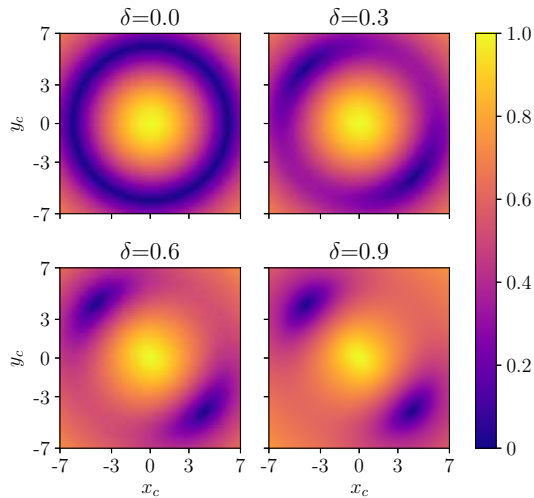}
	\caption{(Color online) Tomography of absolute value of correlation, $g^{(1)}(\vb{r}_c)$ with $ \vb{r}_c= x_c \vu*{x} + y_c \vu*{y} $, at z=0 plane of BEC for different values of shift parameters with $\beta=\pi/4$ of the vortex beam.}
	\label{fig:cor_abs}
\end{figure}

\begin{figure}[H]
	\centering
	\includegraphics[width=0.5\textwidth]{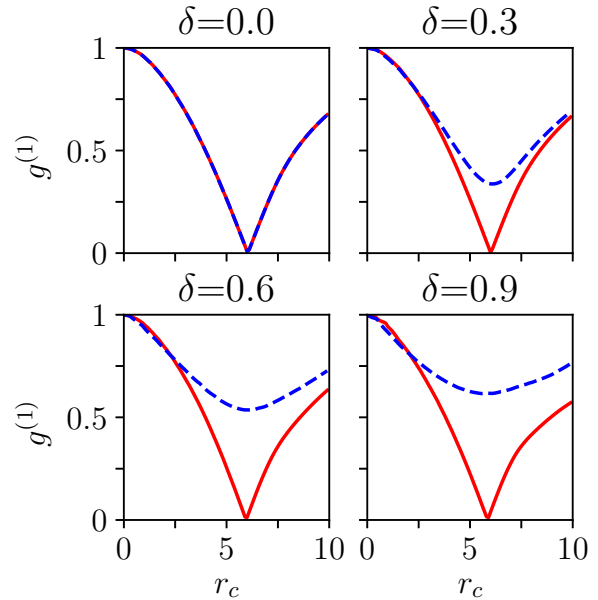}
	\caption{(Color online) Variation of absolute value of correlation, $g^{(1)}(\vb{r}_c)$ (FIG. \ref{fig:cor_abs}), w.r.t. radial distance from the trap center of BEC along the lines $x_c=y_c$ (blue dashed line) and $x_c=-y_c$ (red line) for different values of shift parameter $\delta$ of the vortex beam.}
	\label{fig:cor_45}
\end{figure}

Here we have considered correlation for different ordering scales over the condensate surface using  surface integral. FIG. \ref{fig:cor_abs} demonstrates how the absolute value of  first order correlations or degree of coherence vary with spatial order $ \vb{r}_c= x_c \vu*{x} + y_c \vu*{y} $ for different value of the shift parameter, $\delta$,  of the aLG beam in the case of $\beta=\pi/4$, here $ \vu*{x} \text{ and } \vu*{y} $ are the unit vectors along the transverse coordinates of the trap. The tomography of correlation order shows smaller spatial ordering for matter vortex, which is otherwise long range  \cite{leeuw_2014_phase}  and axisymmetric. Another noticeable point is  directional specific coherent structure of spatial ordering and this differentiability from the axisymmetry becomes larger with the increased values of $\delta$. 
Moreover, this is consistent with the density profile shown in FIG. \ref{fig:bec_vortex_super}. To quantify the variations of coherence, we plot the absolute value of $g^{(1)}$ at FIG. \ref{fig:cor_45} with respect to spatial order  $ r_c (= \sqrt{x_c^2+y_c^2})$ along the lines $x_c=y_c$ (blue dashed line) and $x_c=-y_c$ (red line) on the $ z=0 $ plane. These directions are chosen as we find large variation of $g^{(1)}$ values along the line $x=y$ with respect to $\delta$. Along both the lines, $g^{(1)}$ is minimum nearly at $ r_c=6 $. Again, this direction of large variation of degree of coherence changes with the value of $\beta$, the phase of shifting parameter of the aLG beam. These datum of coherence will be useful ingredient for atom interferometry experiments  \cite{pasquini_2005_atom} using condensed atoms.

\section{Conclusion}
In this paper, we have derived a theory of interaction for dipole and quadrupole transitions in atomic BEC due to the external field of aLG beam. The asymmetric property in the beam has been incorporated by   considering a complex-valued shift to a conventional LG beam in the Cartesian plane. We have shown that  aLG beam, where the vortex center coincides with the beam axis, can be considered as a weighted superposition of an infinite number of LG modes having consecutive topological charges of increasing order with same orientation as the unshifted LG beam. Transfer of multiple OAM from the beam to the BEC generate a superposition of vortex states sharing same intermediate electronic state in two-photon Raman transition. We have found that the dipole and quadrupole Rabi frequencies as well as the superposition of the vortex states can be controlled externally by changing the asymmetry parameter of the beam. This asymmetric effects of the aLG beam could be experimentally corroborated  by measuring the OAM in the  BEC using surface wave
spectroscopy \cite{chevy_2000_measure,haljan_2001_use}.
The fascinating phenomenon of producing multiply quantized vortices in BEC has been the subject of intensive research in superconductivity \cite{milorad_2015_emergent}, superfluid Fermi gases \cite{zwierlein_2005_vortices,zwierlein_2006_fermionic,zwierlein_2006_direct} and even in real material \cite{chmiel_2018_observation}. A significant enhancement of quadrupole Rabi frequency has been observed for certain complex-valued shift in the beam. 
Moreover, the study of degree of coherence of the final condensed state shows interesting directional dependent variations at the $z=0$ plane of the trap, which may lead to interesting physics in atom interferometry.\\
There are several research directions to consider in future efforts. One straight forward would be to study how the BEC evolves dynamically while interacting with the aLG beam. Additionally, one could investigate the generation of spin orbit coupling in ultra cold gases by using aLG beam. Also, such study could be extended to the cases where long range interaction such dipolar is present in the BEC.

\appendix*
\section{}\label{ap:calc_step}
In Eq. \eqref{eq:alg}, putting $ s^2= \rho^2 - 2\rho\delta e^{\im \beta} e^{\im \varphi} $, we get
\begin{align*}
\mathcal{U}'_{l}(\rho,\varphi,z)&=\mathcal{A}~\mathcal{U}_{l}(\rho,\varphi,z) \times \nonumber \\ &\exp[2\rho\delta e^{\im \beta} e^{\im \phi} \qty(\frac{1}{w^2(z)}-\frac{\im k z}{2(z^2+z_R^2)})] \nonumber\\
\end{align*}
Now,
\begin{align*}
\frac{1}{w^2(z)}-\frac{\im k z}{2(z^2+z_R^2)}=\frac{z_R^2}{w_0^2(z^2+z_R^2)}-\frac{\im k z}{2(z^2+z_R^2)}\\
=\frac{2z_R^2-\im k w_0^2 z}{2w_0^2(z^2+z_R^2)}
=\frac{z_R^2-\im z_R z}{w_0^2(z^2+z_R^2)}
=\frac{z_R}{w_0^2(z_R+\im z)}\\
=\frac{1}{w_0 w(z)}e^{-\im \zeta(z)}
\end{align*}
So,
\begin{align*}
\mathcal{U}'_{l}(\rho,\varphi,z)&=\mathcal{A}~\mathcal{U}_{l}(\rho,\varphi,z)\exp[\frac{2\rho\delta e^{\im \beta} e^{\im \phi} }{w_0 w(z)}e^{-\im \zeta(z)}]
\end{align*}
\bibliographystyle{apsrev4-1}
\bibliography{reference}
\end{document}